\documentstyle[11pt,psfig]{article}
\begin{document}

\centerline{{\large\bf Under embargo at NATURE (accepted July 27; submitted
June 11, 1998)}}

\bigskip \bigskip 
\centerline{{\large\bf A `Hypernova' model for SN 1998bw associated
with gamma-ray burst of 25 April 1998}
\footnote{Partially based on observations collected at ESO-La Silla} }

\bigskip

\bigskip
\centerline{\large K.~Iwamoto$^\ast$, P.A.~Mazzali$^\dagger$,
K.~Nomoto$^{\ast,\dagger\dagger}$, H.~Umeda$^{\ast, \dagger\dagger}$,
T.~Nakamura$^\ast$,}
\centerline{\large F.~Patat$^\ddagger$, I.J.~Danziger$^\dagger$,
T.R.~Young$^\ast$, T.~Suzuki$^{\ast, \dagger\dagger}$,
T.~Shigeyama$^{\ast, \dagger\dagger}$,}
\centerline{\large T.~Augusteijn$^\ddagger$, V.~Doublier$^\ddagger$,
J.-F.~Gonzalez$^\ddagger$, H.~Boehnhardt$^\ddagger$,
J.~Brewer$^\ddagger$} 
\centerline{\large O.R.~Hainaut$^\ddagger$, C.~Lidman$^\ddagger$,
B.~Leibundgut$^\star$, E.~Cappellaro$^\S$, M.~Turatto$^\S$,}
\centerline{\large T.J.~Galama$^\parallel$,
P.M.~Vreeswijk$^\parallel$, C.~Kouveliotou$^\P$, J.~van
Paradijs$^{\parallel, \ddagger\ddagger}$,}
\centerline{\large E.~Pian$^{\ast\ast}$, E.~Palazzi$^{\ast\ast}$,
F.~Frontera$^{\ast\ast}$}

\bigskip

\centerline{$^\ast$Department of Astronomy, School of Science,
University of Tokyo, Tokyo 113-0033, Japan}

\centerline{$^\dagger$Osservatorio Astronomico di Trieste, via G.B.
Tiepolo 11, I-34131 Trieste, Italy}

\centerline{$^\ddagger$European Southern Observatory, Casilla 19001,
Santiago 19, Chile}

\centerline{$^\S$Osservatorio Astronomico di Padova, 
 vicolo dell'Osservatorio 5, I-35122 Padova, Italy}

\centerline{$^\parallel$Astronomical Institute ``Anton Pannekoek'',
University of Amsterdam, }

\centerline{and Center for High Energy Astrophysics, Kruislaan 403,
1098 SJ Amsterdam, The Netherlands}

\centerline{$^\star$European Southern Observatory,
Karl-Schwarzschild-Strasse 2, D-85748, Garching, Germany}

\centerline{$^\P$NASA Marshall Space Flight Center, ES-84, Huntsville,
AL 35812, USA}

\centerline{$^{\ast\ast}$Istituto Tecnologie e Studio Radiazioni 
Extraterrestri, CNR, Bologna, Italy}

\centerline{$^{\dagger\dagger}$Research Center for the Early Universe,
School of Science, University of Tokyo, Tokyo 113-0033, Japan}

\centerline{$^{\ddagger\ddagger}$Department of Physics, University of
Alabama, Huntsville, AL 35899, USA}

\newpage


\bigskip

{\noindent\bf The discovery of the peculiar supernova (SN) 1998bw and
its possible association with the gamma-ray burst (GRB)
980425$^{1,2,3}$ provide new clues to the understanding of the
explosion mechanism of very massive stars and to the origin of some
classes of gamma-ray bursts.  Its spectra indicate that SN~1998bw is a
type Ic supernova$^{3,4}$, but its peak luminosity is unusually high
compared with typical type Ic supernovae$^3$.  Here we report our
findings that the optical spectra and the light curve of SN 1998bw can
be well reproduced by an extremely energetic explosion of a massive
carbon+oxygen (C+O) star.  The kinetic energy is as large as $\sim 2-5
\times 10^{52}$ ergs, more than ten times the previously known energy
of supernovae. For this reason, the explosion may be called a
`hypernova'. Such a C+O star is the stripped core of a very massive
star that has lost its H and He envelopes. The extremely large energy,
suggesting the existence of a new mechanism of massive star explosion,
can cause a relativistic shock that may be linked to the gamma-ray
burst.}

\bigskip

SN 1998bw is classified spectroscopically as a type Ic supernova,
because its optical spectra lack any hydrogen and helium features and
the Si II absorption feature is very different from those of type Ia
supernovae $^{5}$.  Two recent type Ic supernovae, SNe 1994I$^{6,7}$
and 1997ef$^{8}$, have somewhat similar spectra to that of SN 1998bw
and their light curves were well reproduced by models of the
collapse-induced explosion of C+O stars~(Fig.1).  This has led us to
construct hydrodynamical models of exploding C+O stars also for SN
1998bw.  The model parameters are the stellar mass $M_{\rm CO}$, the
explosion energy $E_{\rm exp}$, and the mass of the synthesized
$^{56}$Ni $M_{56}$, assuming that the light is generated by the
$^{56}$Ni decay as in type Ia supernovae.

Despite their spectral similarity, these three type Ic supernovae have
distinctly different brightnesses and light curve shapes as seen in
Figure 1.  This is because the brightness and light curve shape depend
mainly on $M_{56}$ and on the pair of values ($E_{\rm exp}$, $M_{\rm
CO}$), respectively, while the spectral features are sensitive to the
chemical composition, which is basically similar in the C+O stars.
The peak bolometric luminosity $\sim 1.6 \times 10^{43}$ ergs s$^{-1}$
implies that SN 1998bw produced $\sim 0.7 M_\odot$ of $^{56}$Ni, which
is much more than in SNe 1994I$^{6,7}$ and 1997ef$^{8}$.

The above parameters are tightly constrained by comparing the light
curves (Fig. 1), synthetic spectra (Fig. 2), and photospheric
velocities (Fig. 3) with the observations of SN 1998bw$^{3}$.  We find
that the optical properties of SN~1998bw are best reproduced by a
model with $M_{\rm CO} = 13.8 M_\odot$, $E_{\rm exp} = 3 \times
10^{52}$ ergs, and $M_{56} = 0.7 M_{\odot}$ (hereafter designated as
CO138). A C+O star of this mass originates from a $\sim$ 40
$M_{\odot}$ main sequence star.  A compact remnant of mass $M_{\rm
rem} = 2.9 M_{\odot}$ must have been left behind for $0.7 M_{\odot}$
$^{56}$Ni to be ejected, as required to reproduce the brightness of
SN~1998bw.  $M_{\rm rem}$ exceeds the upper mass limit for a stable
neutron star, suggesting the formation of a black hole.

In order to reproduce the light curve of SN 1998bw, the time of core
collapse should be set to coincide with the detection of GRB 980425 to
within +0.7/-2 days.  The rapid rise of the light curve requires the
presence of radioactive $^{56}$Ni near the surface, implying that
large-scale mixing of material took place because of hydrodynamical
instabilities.  The light curve shape can be reproduced with different
explosion models, because the peak width $\tau_{\rm LC}$, which
reflects the time scale of photon diffusion, scales approximately as
$\tau_{\rm LC} \propto \kappa^{1/2} M_{\rm ej}^{3/4} E_{\rm
exp}^{-1/4}$ (ref.13), where $M_{\rm ej} = M_{\rm CO}-M_{\rm rem}$ is
the mass of the ejected matter, and \(\kappa\) denotes the optical
opacity.  However, the photospheric velocity scales in a different
manner, as $ v \propto M_{\rm ej}^{-1/2} E_{\rm exp}^{1/2}$, so that
both $M_{\rm ej}$ and $E_{\rm exp}$ can be constrained from the
spectroscopic data as follows.

Synthetic spectra$^{14}$ of various explosion models were compared
with the observed spectra of SN~1998bw at 3 epochs: May 3, 11 and 23.
The observed featureless spectra are the result of the blending of
many metal lines reaching large velocity and with a large velocity
spread.  Extensive blending can only be achieved with models that have
a large mass in high velocity regions. Therefore, the models that are
more massive and have a larger kinetic energy give better fits.  For
models with $ M_{\rm ej} < 10 M_\odot$, the photosphere forms at
velocities much smaller than those of the observed lines and the lines
do not blend as much as in the observed spectra.  Figure 2 shows that
model CO138 gives consistent fits to the spectra at all three epochs.

Figure 3 shows that the evolution of the photospheric velocity
computed from model CO138 (appeared in {\it lines}) agrees with that
obtained from spectral fits ({\it filled circles}), and with the
observed velocities of the Si II ({\it open circles}) and Ca II ({\it
square}) lines, within the uncertainty arising from the light curve
fitting ({\it dotted lines}). These velocities are among the highest
ever measured in supernovae of any types and thus the smaller mass C+O
star progenitors can be ruled out. By taking into account the
uncertainties, we conclude that massive C+O star models with $E_{\rm
exp} \sim 2-5 \times 10^{52}$ ergs and $M_{\rm CO} \sim 12 - 15
M_\odot$ reproduce the observed light curve and spectra of SN 1998bw
well.

Here we call the supernova with such an extremely large explosion
energy ( $> 10^{52}$ ergs) a `hypernova'$^{15}$ .  The evolutionary
process leading to the hypernova could be as follows: The massive
progenitor of initially $\sim 40 M_\odot$ had a particularly large
angular momentum and a strong magnetic field owing possibly to the
spiraling-in of a companion star in a binary system.  The collapse of
the massive Fe core at the end of the evolution led to the formation
of a rapidly rotating black hole.  Then the large rotational energy of
the black hole was extracted with a strong magnetic field to induce a
successful explosion$^{15,16}$.

The hypernova could induce a gamma-ray burst in the following way: At
the shock breakout in the energetic explosion, the surface layer is
easily accelerated to produce a relativistic shock. When it collides
with circumstellar or interstellar matter, non-thermal electrons that
are produced at the shock front emit high-energy photons via
synchrotron emission.  The energy of these photons is given by $\sim$
160 keV $(\Gamma /100) n_1^{1/2}$(ref.17), where $\Gamma$ denotes the
Lorentz factor of the expanding shell and $n_1$ is the density of the
interstellar matter in cm$^{-3}$.  Thus the event could be observed as
a gamma-ray burst if $\Gamma$ becomes as large as $\sim$ 100.  Our
preliminary calculations show that spherically symmetric models may
not produce large enough energies in gamma-rays.  However, an
axi-symmetric explosion could produce particularly high speed material
by a focused shock wave in the polar direction.  The strong radio
emission at early phases, which suggests the existence of such a
relativistic flow$^{18}$, is consistent with the above scenario.
Preliminary spectral polarization measurements show that polarization
is small ($\sim 1$\% but possibly decreasing between 4 May and 20
May). Some degree of asymmetry in the envelope morphology is therefore
possible, but the precise form depends on the undetermined orientation
relative to the line-of-sight.  In the near future, late time spectra
will provide the heavy element abundances and their velocities in SN
1998bw to test our prediction (given in the legend of Figure 1). The
late time decline rate of the light curve is also expected to give
further constraints on the model parameters$^{19}$.

\vspace{3cm}

\noindent
------------------------------------------------------------------------------------------------------------------------------------

\noindent
1. Soffitta,~P., Feroci,~M., \& Piro,~L. {\sl IAU Circ.} No. 6884 (1998).


\noindent
2. Lidman, C., Doublier, V., Gonzalez, J.-F., Augusteijn, T., Hainaut,
O.R., Boehnhardt, H., Patat, F., Leibundgut, B. {\sl IAU Circ.} 
No. 6895 (1998).

\noindent
3. Galama, ~T.J. {\sl et al.}  
Discovery of the peculiar supernova SN 1998bw in the error box of
the \(\gamma\)-ray burst of 25 April 1998.
{\sl Nature} (in press); astro-ph/9806175 (1998) 

\noindent
4. Patat, ~F., \& Piemonte,~A. {\sl IAU Circ.} No. 6918 (1998).

\noindent
5. Filippenko, ~A.~V. 
Optical Spectra of Supernovae.
{\sl Annu. Rev. Astron. Astrophys.} {\bf 35}, 309-355 (1997).

\noindent
6. Nomoto, K., Yamaoka, H., Pols, O.R., Van den Heuvel, E.P.J.,
Iwamoto, K., Kumagai, S., Shigeyama, T.
A carbon-oxygen star as progenitor of the type Ic supernova
1994I.
{\sl Nature} {\bf 371} 227-229 (1994).

\noindent
7. Iwamoto, K., Nomoto, K., H\"oflich, P., Yamaoka, H., Kumagai, S.,
\& Shigeyama, T. 
Theoretical Light Curves for the Type Ic Supernova SN 1994I. 
{\sl Astrophys. J.} {\bf 437} L115-L118 (1994).

\noindent
8. Iwamoto, ~K., Nakamura,~T., Nomoto,~K., Mazzali,~P.A.,
Garnavich,~P., Kirshner,~R., Jha,~S., \& Balam, ~D. 
Light curve modeling of the type Ib/Ic supernova 1997ef.
{\sl Astrophys. J.} (submitted); astro-ph/9807060 (1998).

\noindent
9. Garnavich, ~P., Jha, ~S., Kirshner, ~R., \& Challis, ~P.  {\sl IAU
Circ.} No. 6798 (1997).

\noindent
10. Richmond, ~M.~W., {\em et al.} {\sl UBVRI} photometry of the Type
 Ic SN 1994I in M51. {\sl Astronomical J.} {\bf 111}, 327-339 (1996).

\noindent
11. Thielemann, ~F.,~-K., Nomoto, K., \& Hashimoto, M.  
Core-Collapse Supernovae and Their Ejecta.
{\sl Astrophys. J.} {\bf 460} 408-436 (1996).

\noindent
12. Nakamura, ~T., Umeda,~T., Nomoto,~K., Thielemann,~F.,~-K., \&
Burrows, A. Nucleosynthesis in type II supernovae and abundances in
 metal poor stars. {\sl Astrophys. J.} (submitted) (1998).

\noindent 
13. Arnett,~W.D. Type I supernovae.I.Analytic solutions for the early
part of the light curve. {\sl Astron. Astrophys} {\bf 253} 785-797
(1982).

\noindent
14. Mazzali,~P.A., Lucy,~L.B.  
The application of Monte Carlo methods to the synthesis
    of early-time supernovae spectra.
{\sl Astron. Astrophys}, {\bf 279} 447-456 (1993). 

\noindent
15. Paczy\'nski, ~B. Are Gamma-Ray Bursts in Star-Forming Regions?
{\sl Astrophys. J.} {\bf 494,} L45--L48 (1998).

\noindent
16. Woosley, ~S.~E. Gamma-ray bursts from stellar mass accretion disks
  around black holes. {\sl Astrophys. J.} {\bf 405,} 273--277 (1993).

\noindent
17. Piran,~T.
Towards Understanding Gamma-Ray Bursts.
in {\sl Some Unsolved Problems in Astrophysics} (eds. J. N. Bahcall \&
J. P. Osriker) 343--377 (Princeton Univ. Press, Princeton, 1997).

\noindent
18. Kulkarni, ~S.~R., Bloom,~J.~S., Frail,~D.~A., Ekers,~R.,
Wieringa,~M., Wark, ~R., \& Higdon, ~J.~L. {\sl IAU Circ.} No. 6903
(1998).

\noindent
19. Baron, ~E., Young, ~T.~R., Branch, ~D. {\sl Astrophys. J.} {\bf
409,} 417--421 (1993).

\noindent
------------------------------------------------------------------------------------------------------------------------------------

\newpage

\begin{figure}[t]
\centerline{\psfig{figure=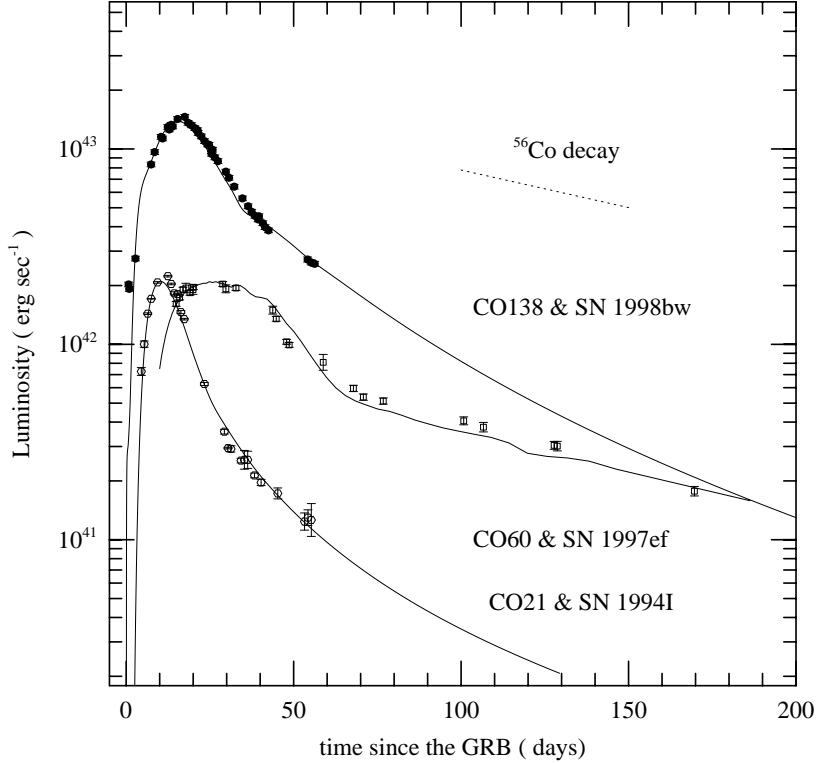,width=13.3cm}}
\vspace{-2.8cm}
\caption[98bwaph.fig1.ps]{\small
The bolometric light curve of model CO138 ($M_{\rm CO} = 13.8
M_\odot$, $E_{\rm exp} = 3 \times 10^{52}$ ergs, $M_{56} = 0.7
M_\odot$) compared with the observations of SN1998bw. The time of the
core collapse is set at the detection of the GRB980425. The distance
to the host galaxy ESO 184-G82 is taken to be $\sim 39 \pm 1$ Mpc, as
estimated from the redshift $z \sim 0.0085 \pm 0.0002$ and a Hubble
constant 65 km s$^{-1}$ Mpc$^{-1}$.  The light curves of other type Ic
supernovae SNe 1997ef$^{8,9}$ and 1994I$^{10}$ are also shown, for
comparison, together with the corresponding theoretical models, CO60
(ref.8, $M_{\rm CO} = 6.0 M_\odot$, $E_{\rm exp} = 10^{51}$ ergs,
$M_{56} = 0.15 M_\odot$) and CO21 (ref.6, $M_{\rm CO} = 2.1 M_\odot$,
$E_{\rm exp} = 10^{51}$ ergs, $M_{56} = 0.07 M_\odot$), respectively.
Note that $E_{\rm exp}$ and $M_{56}$ of CO138(SN 1998bw) are very much
larger than in the models for the other two type Ic supernovae.  The
observed V light curves are transformed into the bolometric light
curves assuming that the bolometric correction is negligible.  The light
curves are computed with a radiative transfer code$^{8}$, assuming a
detailed balance between photo-ionizations and recombinations and
adopting a simplified treatment of line opacity.  The explosive
nucleosynthesis was calculated using a detailed nuclear reaction
network$^{11,12}$ including a total of 211 isotopes up to
$^{71}$Ge. Our calculation predicts the amount of other radioactive
nuclei, 1.4 $\times 10^{-3}$ M$_\odot$ $^{44}$Ti and 1.4 $\times
10^{-2}$ M$_\odot$ $^{57}$Ni, and other stable elements, $^{16}$O:
7.6, $^{20}$Ne: 0.44, $^{23}$Na: 1.2 $\times 10^{-6}$, $^{24}$Mg:
0.46, $^{27}$Al: 0.18, $^{20}$Si: 0.82, $^{40}$Ca: 5.0 $\times
10^{-2}$, $^{20}$Ne: 0.44 (in $M_\odot$).
}

\end{figure}

\newpage

\begin{figure}[t]
\centerline{\psfig{figure=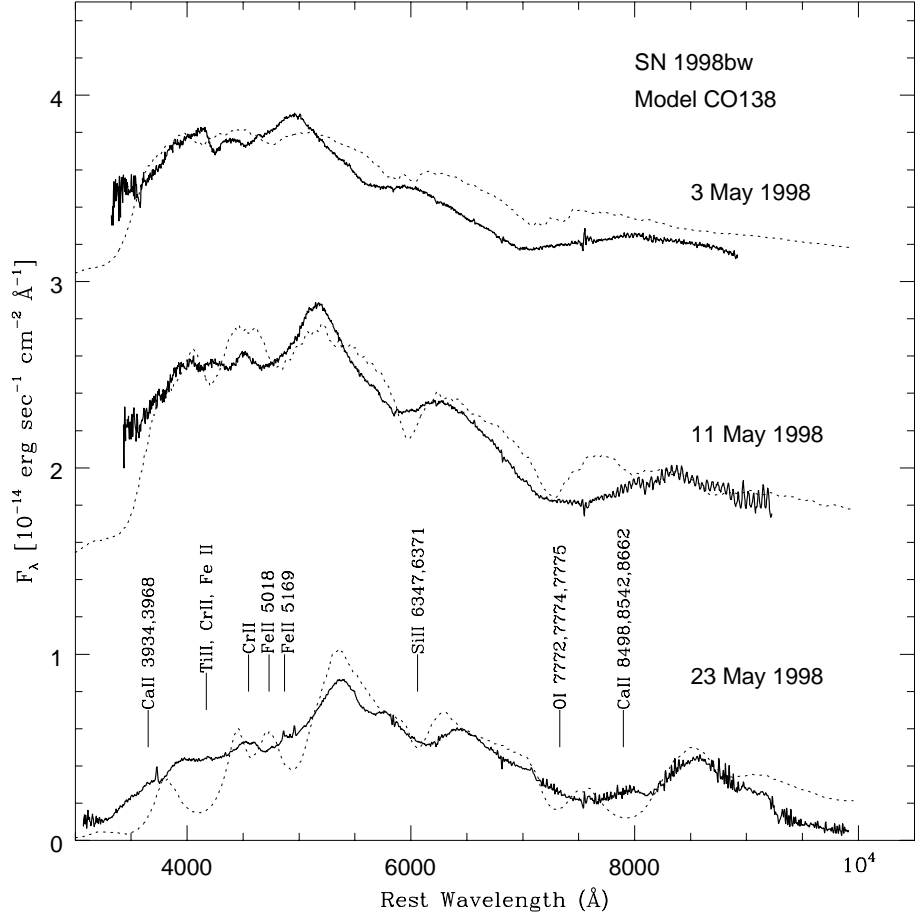,width=13.3cm}}
\caption[98bwaph.fig2.ps]{\small
Three observed spectra (full lines, Patat {\sl et al.}, in
preparation), where the galaxy background has been subtracted, are
compared with the synthetic spectra (dashed lines) computed with the
Monte Carlo code$^{14}$, improved with the inclusion of photon
branching (Mazzali \& Lucy, in preparation), using model CO138. The
synthetic spectra were computed using the luminosity derived from the
light curve and a distance of 39 Mpc, and assuming no reddening. The
observed featureless spectra are the result of the blending of many
metal lines reaching large velocity and with a large velocity
spread. The apparent emission peaks are actually low opacity regions
of the spectrum where photons can escape. The 3 May and the 11 May
spectra have been shifted upwards by 3.0 and $1.5 \times 10^{-14}$ erg
s$^{-1}$ cm$^{-2}$ \AA$^{-1}$, respectively.  The most important lines
are marked on the 23 May spectrum, but they also contribute to the 3
May and 11 May spectra, although with somewhat different ratios. Line
blending in the case of SN~1998bw is even more severe than it was in
the massive type Ic supernova 1997ef$^{8}$, indicating an even larger
mass.  The massive progenitor model is the only one that gives the
correct extent of line blending.  Differences in the blue band between
the observed spectrum and the synthetic one are probably due to
uncertainties in the determination of the abundance and distribution
of Fe-group elements in high velocity parts of the ejecta.  The
possible presence of O I line absorption in the early spectra
complicates any derivation of velocities in the high velocity wings of
the feature conceivably ascribed to Ca II absorption.
}

\end{figure}

\newpage

\begin{figure}[t]
\centerline{\psfig{figure=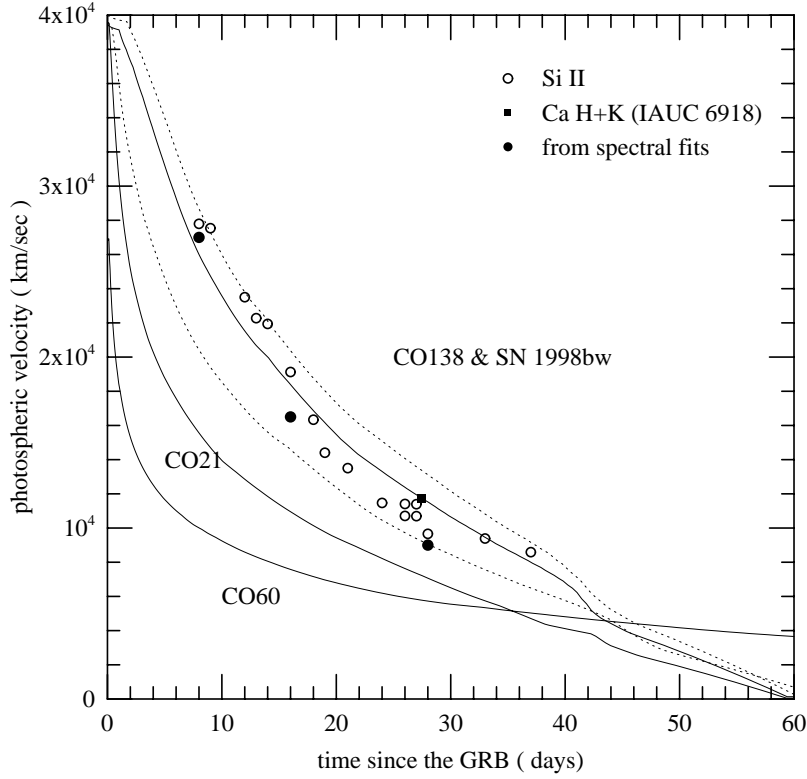,width=13.3cm}}
\caption[98bwaph.fig3.ps]{\small
The evolution of the calculated photospheric velocities of CO138,
 CO60, and CO21 (solid lines), the photospheric velocities obtained
 from spectral models (filled circles), the observed velocity of the
 Si II 634.7, 637.1 nm line measured in the spectra at the absorption
 core (open circles, Patat {\sl et al.} in preparation), and that of
 the Ca II H+K doublet measured in the spectrum of May 23 (square,
 ref.4).  The observed velocities are in good agreement with the
 photospheric velocities of CO138, which are much larger velocity than
 in CO21 and CO60 because of the much larger explosion energy. The
 upper and lower dotted lines are the velocities of models with
 ($M_{\rm CO}$, $E_{\rm exp}$) = (15 $M_{\odot}$, 5 $\times 10^{52}$
 ergs) and (12 $M_{\odot}$, 2 $\times 10^{52}$ ergs), respectively.
 The light curves of these two models also fit SN 1998bw well.  This
 indicates the acceptable ranges of $M_{\rm CO} \sim 12 - 15
 M_{\odot}$ and $E_{\rm exp} \sim 2 -5 \times 10^{52} $ ergs.
}

\end{figure}

\end{document}